\documentclass[aps,pre,twocolumn,superscriptaddress,showpacs,floatfix]{revtex4}%

\usepackage{epsfig}
\usepackage{times}
\usepackage{amssymb}
\usepackage{colordvi}
\usepackage{color}     
\usepackage{amsbsy}
\usepackage{amsmath}
\usepackage{amsbsy}
\usepackage{latexsym}
\usepackage{graphicx}

{\rm \addcolor $\bullet|$ \blackcolor}                

 \newcommand{\beas}[1]{\begin{subequations}\label{#1}\bea}
 \newcommand{\eeas}{\eea\end{subequations}}

 \newcommand{\bbf}[1]{\mbox{\boldmath{$#1$}}}

 \renewcommand{\r}{{\bf r}}

 \newcommand{\mbf}[1]{{\bf #1}}
 
 \newcommand{\x}{\mbf{x}}
 \newcommand{\e}{\mbf{e}}
 \newcommand{\M}{\mbf{M}}
 \renewcommand{\r}{\mbf{r}}
 \newcommand{\p}{\mbf{p}}

 \newcommand{\bPi}{{\mbf{\Pi}}}
 
 \newcommand{\bLambda}{\mbf{\Lambda}}
 \newcommand{\bkappa}{\boldsymbol{\kappa}}
 \newcommand{\bv}{\mathbf{v}}
 \newcommand{\z}{\mbf{z}}

 \newcommand{\Nch}{N_{\rm ch}}
 
 \newcommand{\kB}{k_{\rm B}}
 
 \newcommand{\taus}{\tau_s}
 \newcommand{\feq}{f_\mathrm{eq}}
 
 \newcommand{\avee}[1]{\langle #1\rangle}
 
 \newcommand{\ave}[1]{\left\langle #1\right\rangle}
 
 \newcommand{\be}{\begin{equation}}
 \newcommand{\ee}{\end{equation}}
 \newcommand{\bea}{\begin{eqnarray}}
 \newcommand{\eea}{\end{eqnarray}}

 \newcommand{\bsigma}{\boldsymbol{\sigma}}

 \newcommand{\stress}{\boldsymbol{\sigma}}

 \newcommand{\Ntot}{N_{\rm b}}

 \newcommand{\DeltaLambda}{\delta\bLambda}

 \newcommand{\QM}[1]{\Blue{{\bf (}#1{\bf )}}}
 \newcommand{\QMn}[1]{(#1)}

 \newcommand{\Qs}{\QM{i} }
 \newcommand{\QQs}{\QM{ii} }
 \newcommand{\QQQs}{\QM{iii} }
 \newcommand{\QQQQs}{\QM{iv} }
 \newcommand{\QQQQQs}{\QM{v} }
 \newcommand{\Qn}{\QMn{i}}
 \newcommand{\QQn}{\QMn{ii}}
 \newcommand{\QQQn}{\QMn{iii}}
 
 \newcommand{\QQQQQn}{\QMn{v}}
 \newcommand{\Qns}{\QMn{i} }

 \newcommand{\QQQQQns}{\QMn{v} }

\begin{document}

\title{Systematic time-scale-bridging molecular dynamics applied to flowing polymer melts}

\author{Patrick Ilg, Hans Christian \"Ottinger, and Martin Kr\"oger}
\affiliation{Polymer Physics, ETH Z\"urich, Department of Materials, CH-8093 Z\"urich, Switzerland}


\begin{abstract}
We present a novel thermodynamically guided, low-noise, time-scale
bridging, and pertinently efficient strategy for the dynamic
simulation of microscopic models for complex fluids. The systematic
coarse-graining method is exemplified for low-molecular polymeric
systems subjected to homogeneous flow fields. We use established
concepts of nonequilibrium thermodynamics and an alternating
Monte-Carlo--molecular dynamics iteration scheme in order to obtain
the model equations for the slow variables. For chosen flow
situations of interest, the established model predicts structural as
well as material functions beyond the regime of linear response. As
a by-product, we present the first steady state equibiaxial
simulation results for polymer melts. The method is simple to
implement and allows for the calculation of time-dependent behavior
through quantities readily available from the nonequilibrium steady states.
\end{abstract}

\maketitle

\section{Introduction}
Systematic bridging the time- and length-scale gap between
microscopic and macroscopic levels of description is ``of the
greatest importance in theoretical science'' \cite{gellmann2007}. In
many cases, this challenging task can neither be solved purely
analytically nor by brute force computer simulations alone. This is
true in particular for soft condensed matter like colloids,
polymers, liquid crystals, with their internal structure leading to
additional length and time scales, intermediate between microscopic
and macroscopic scales \cite{larsonbook}.

In recent years, effective interactions for coarse-grained models of
soft matter systems have been derived from inversion procedures that
are designed to reproduce chosen pair correlation functions
\cite{krakoviackCGcopolym,kleinCGreview,coarsegrained2000,Kremer_reviewMultiscale,guenza_reviewCG}.
While the inversion procedures often reproduce the static structure
rather accurately, their naive extension to dynamical phenomena
clearly failed \cite{kleinCGreview}. This deficiency calls for a
systematic approach that bridges simultaneously the time- and
length-scale gap between two levels.
For comparatively simple two-dimensional crystalline solids, a
simultaneous space/time coarse-graining procedure was proposed
recently in \cite{curtarolo} based on renormalization group
techniques. There, temporal coarse graining is coupled via the
dynamical critical exponent to the degree of spatial coarse
graining.
This approach is unfortunately not applicable to the dynamics of
complex fluids, since their internal structures break the scale
invariance - an essential prerequisite for renormalization group
methods - and lead to the emergence of slow, non-hydrodynamic modes.
The latter are typically described on an intermediate, mesoscopic
level by a set of ``collective'' or ``structural'' variables
$\bPi(\z)$ which in turn determine the macroscopic properties of
complex fluids \cite{larsonbook}. Since many microstates $\z$ are
compatible with given values of $\bPi$, the mesoscopic level is
necessarily stochastic in nature. Thus, the emergence of entropy and
irreversibility from reversible dynamics is the hallmark of coarse
graining.
Several coarse-graining approaches, in particular for solutions and
suspensions, have been suggested where the starting level is already
dissipative (see e.g.~\cite{guenza_reviewCG,Larson_CGchain} and
references therein). In the context of polymer melts, promising work
on coarse-graining polymer chains starting from Hamiltonian dynamics
has been done e.g.~in \cite{Briels_CGchain,BrielsPadding}.

In this paper, we propose and explore a systematic,
thermodynamically guided method which establishes the mesoscopic
model from the underlying microscopic level. The proposed method is
general enough to be applied to various soft matter systems and
valid in equilibrium as well as nonequilibrium situations. Its
strategy relies on the balance of reversible and irreversible
contributions to the dynamics and explicitly accounts for the
entropy generated in the coarse-graining step \cite{hco_lessons}. We
use an alternating Monte-Carlo (MC) and molecular dynamics (MD)
simulation scheme in order to iteratively determine static and
dynamic ``building blocks'' \cite{hcobook2} of the mesoscopic model
self-consistently.

\begin{figure}[tb]
\begin{center}
\includegraphics[width=9cm, angle=0]{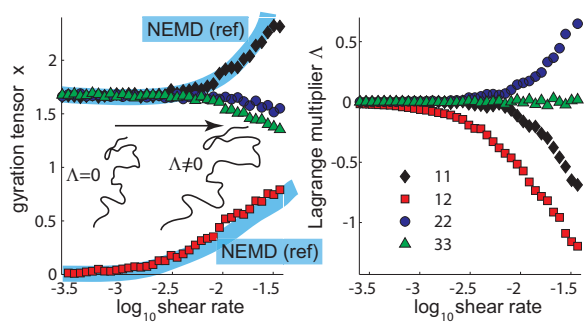}
\end{center}
\caption{(color online). Components of gyration tensor $\x$ (left) and Lagrange multiplier $\bLambda$
(right panel) vs.~shear rate for a FENE polymer melt ($N=20$).
Lagrange multipliers self-consistently enter the anisotropy and stretching of polymer chains.
Comparison with standard NEMD reference results (left panel) show that the generalized
canonical distribution (\ref{newrhoz}) provides a good description of the nonequilibrium
stationary state in shear flow.
We use Lennard-Jones units throughout this paper.}
\label{coarse_grain_2008_Fig1}
\end{figure}

\section{Original and Coarse-Grained Model System}
The novel algorithm is applicable to a wide range of soft matter systems.
In order to illustrate the basic idea and its worked out counterpart, let us consider
 a particular liquid, a classical monodisperse bulk model polymer
 melt.
 The system consists of $\Nch$ anharmonic multibead-spring (FENE) chains made of $N$ purely repulsive Lennard-Jones
 beads each \cite{hoy,mkbook,daivis}; $\Ntot=\Nch N$ particle positions and momenta are denoted as
 $\{\r_j\}$ and $\{\p_j\}$. The interaction energy between particle $i$ and $j$
is $U_{ij}=U^{\rm LJ}_{ij}+U^{\rm FENE}_{ij}$, where
\begin{equation}
 U^\textrm{LJ}_{ij} = 4\epsilon\left[ \left(\frac{\sigma}{r_{ij}}\right)^{12}-\left(\frac{\sigma}{r_{ij}}\right)^{6}+\frac{1}{4} \right]
 \quad \mbox{for }\ r_{ij} \leq 2^{1/6}\sigma,
\end{equation}
and $U^\textrm{LJ}_{ij}=0$ else. The distance between particles $i$
and $j$ is denoted by $r_{ij}$, $\sigma$ the bead diameter and
$\epsilon$ the Lennard-Jones interaction energy. Chain connectivity
is ensured by FENE springs that act between adjacent neighbors along
the chain,
\begin{equation}
 U^\textrm{FENE}_{ij} = -\frac{1}{2}\epsilon_\textrm{FENE}\; \ln\left[ 1 - \left(\frac{r_{ij}}{r_0}\right)^2 \right],
\end{equation}
and $U^{\rm FENE}_{ij}=0$ for all other particle pairs. All model
parameters and thermodynamic state point are adopted from
\cite{mkbook}: temperature $T=\epsilon/\kB$, density
$n=0.84\,\sigma^{-3}$, finite extensibility of the springs
$r_0=1.5\,\sigma$, and the strength of chain potential
$\epsilon_{\rm FENE}=67.5\,\epsilon$ is large enough in order to
prevent chain crossings. In the following, we use reduced
Lennard-Jones units throughout \cite{units}.

The simple FENE model system is very useful to describe the general
dynamical behavior of polymer melts
\cite{mkbook,daivis,hoy,toddreview}. This system serves as our
starting point, providing
 the microscopic (``atomistic'') level of description without any irreversibility built in.
 Under the assumption that the collective variables $\bPi$ capture all relevant physical processes
 on the time scale of interest, the nonequilibrium state of the system is characterized by the
 generalized canonical ensemble,
  \be
  \rho(\z) = \feq(\z) \,e^{-\bLambda:\bPi(\z) - \Lambda_0},
  \label{newrhoz}
 \ee
with phase space coordinates $\z \equiv \{\r_j,\p_j\}$
and the classical $\feq(\z)\propto \exp\{-H(\z)/\kB T\}$ with $H$ denoting the
microscopic Hamiltonian \cite{gellmann2007,hcobook2,ilgcanonical1}.
The Lagrange multipliers $\bLambda(\x)$ (cf.~Fig.~\ref{coarse_grain_2008_Fig1}) are
determined by the values of the slow variables,
$\x=\avee{\bPi(\z)}$,
where the average is performed with (\ref{newrhoz}),
and $\Lambda_0$ a normalization constant.
As structural variable, we here choose $\x$ to be the mean tensor of gyration,
\begin{equation}
 \bPi(\z) = \frac{1}{\Nch N}\sum_{a=1}^{\Nch}\sum_{j=i_a+1}^{i_a+N}(\r_j-\r_c^a)(\r_j-\r_c^a),
\end{equation}
where $i_a=(a-1)N$ and $\r_c^a=N^{-1}\sum_{i=i_a+1}^{i_a+N}\r_i$ is the center of mass
of chain $a$.
This choice of slow variables is appropriate for low-molecular, unentangled polymeric systems,
where $\x$ indeed varies slowly compared with fast relaxation processes such as
fluctuation of bond lengths and angles, intermolecular distances, or higher normal modes
\cite{larsonbook,hcobook2}.
For a more detailed justification of our choice of $\x$ see Appendix \ref{slow.app}.
 We can neglect the macroscopic hydrodynamic velocity field in (\ref{newrhoz})
 since it equilibrates extremely rapidly on length scales of individual polymers \cite{hco1989}.
 The same situation is encountered in other complex fluids
 as long as the {\em large} relaxation time scales of the collective variables
 are generated on relatively {\em short} length scales.
For a more complete treatment including the hydrodynamic fields see
Ref.~\cite{pi_unentangled}.

The time evolution for the slow variables $\x$ can in general be
written as \cite{hcobook2}
 \be
  \dot{\x} = \dot{\x}_{\rm rev} + {\bf M}\,\colon\!\frac{\delta S}{\delta \x}, \qquad
  \frac{\delta S}{\delta \x} = \kB \bLambda,
 \label{xdot}
 \ee
 where $\dot{\x}_{\rm rev}$ denotes the reversible contribution in terms of a Poisson bracket.
 Here, we have employed the expression for the macroscopic entropy
 $S(\x)=-\kB \avee{\ln \rho}$
 corresponding to the ensemble (\ref{newrhoz}).
Entropy gradients drive the irreversible contribution to (\ref{xdot}).
 Equation (\ref{xdot}) is justified e.g.~from projection operator derivation \cite{hcoprojectors,hcobook2},
 which shows that the symmetric friction matrix ${\bf M}(\x)$ can be obtained from
 a Green-Kubo type formula
 \be
 \label{GreenKubo}
  \M = \avee{\bbf{\cal M}(\z(t))}, \quad
  \bbf{\cal M} = \frac{1}{2\kB \taus} \triangle_{\taus}\bbf{\Pi}(\z)\triangle_{\taus}\bbf{\Pi}(\z),
 \ee
 where $\triangle_{\taus}\bbf{\Pi}$ denotes fast fluctuations of $\bPi$ on a time scale
 $\taus$ that separates the evolution of the slow variables $\x$ from the
 rapid dynamics of the remaining degrees of freedom.

The reversible part of motion is obtained analytically
 by considering the transformation behavior of $\bPi$, cf.~\cite{hcobook2} for
 worked out examples.
 Specifically, when $\x$ is a conformation tensor such as the tensor of gyration,
 and considering a macroscopic flow field $\bv(\r)=\bkappa\cdot\r$,
 hence $\bkappa\equiv (\nabla\bv)^T$,
 one finds the so-called upper-convected behavior \cite{pi_unentangled},
$\dot{\x}_{\rm rev}(\x,\bkappa) = \x\cdot\bkappa^T + \bkappa\cdot\x$.
The remaining building blocks $\bLambda$ and ${\bf M}$ needed to
complete the coarse-grained model (\ref{xdot}), we obtain
self-consistently through a hybrid iteration scheme, as described
next.

\section{Systematic Time-Scale Bridging Method}
In general, the space of admissible values for the slow variable $\x$ is too large
 for a full parameterization of $\bLambda(\x)$ from direct numerical integration.
 We choose to parameterize $\bLambda$ and ${\bf M}$ along one-dimensional paths
 $\x(\dot{\gamma})$, where $\dot{\gamma}$ denotes the value of the external control parameter,
 i.e.~the flow rate for chosen velocity gradients $\bkappa(\dot{\gamma})$ in our case.
 Note, that this procedure is analogous to the experimental determination of rheological
 properties in viscometric flows \cite{larsonbook}. 
While errors in determining $\bLambda$ can in principle violate the thermodynamic integrability 
condition for $S(\x)$, this problem is avoided when working with one-dimensional 
paths which do not cross.  
 In order to calculate $\bLambda(\x)$ for relevant $\x$
 (here, relevant for given flow gradient $\bkappa$),
 we investigate nonequilibrium steady states, for which the left hand side
 of (\ref{xdot}) vanishes.
The systematic time-scale bridging method we propose is summarized in Tab.\ \ref{method.table}.

\begin{table}[tb]
\begin{tabular}{cl}
step & \ description \\
\hline\hline
\Qs & \ choose initial values for the Lagrange multipliers $\bLambda$\\
\hline

\QQs & \ generate $n$ independent configurations distributed\\
     & \ according to the generalized canonical ensemble (\ref{newrhoz})\\
\hline

\QQQs & \ solve Hamilton's unconstrained equations of motion\\
      & \ for all $n$ systems during a short time interval $\taus$\\
\hline

\QQQQs & \ calculate the friction matrix ${\bf M}$ from Eq.~(\ref{GreenKubo}) and\\
       & \ $\x$ directly from the $n$ trajectories produced in \QQQn\\
\hline

\QQQQQs & \ calculate an updated value for $\bLambda$ by solving (\ref{xdot})\\
        & \ for $\bLambda$ with $\dot{\x}={\bf 0}$ (in terms of ${\bf M}$, $\x$, and $\bkappa$\\
    & \ the latter two quantities are ``hidden'' in $\dot{\x}_{\rm
    rev}$)\\
\hline\hline
\end{tabular}
\caption{Summary of proposed time-scale bridging method.}
\label{method.table}
\end{table}

The updated Lagrange multipliers obtained in step (v) can potentially be used to re-enter
the procedure at \Qn,
 and follow steps \QQn--\QQQQQns until $\bLambda$ has
 converged.
 The whole procedure (i)--(v) is then repeated for other choices of the control parameter $\dot{\gamma}$
 in order to establish the model (\ref{xdot}) for different external fields.

 Notice, that the strategy does {\em not} require the implementation
 of flow-specific boundary conditions such as Lees-Edwards (shear) \cite{mkbook}
 or Kraynik-Reinelt (planar elongational flow) \cite{toddreview}
 which is a particularly
 useful feature as it allows us to study arbitrary flow situations
 within exactly the same approach. In the same spirit, and in order to
 not potentially falsify results for the friction matrix,
 the algorithm also
 does {\em not} involve any
 constraints such as thermo- or barostats.
 These advantages are build in our approach since the macroscopic variables do not change
 significantly on the short time scale $\taus$ of the MD simulations in \QQQn.

 We now specify how to implement the steps \Qn--\QQQQQns
 efficiently, and how to self-consistently
 determine the range of validity of the underlying assumption (\ref{newrhoz}).
 We choose the control parameter $\dot{\gamma}$
 logarithmically equidistant, $\log(\dot{\gamma})\in[a,a+\Delta a,a+2\Delta a,..,b]$.
 Before we start the procedure, we initialize $\bLambda={\bf 0}$ and $\log \dot{\gamma}=a$.

The loop starts at \Qs with the current value of $\bLambda$. For
\QQs the same $\bLambda$
 is used in a MC scheme to generate microscopic configurations distributed according to (\ref{newrhoz}).
 We have generated $n$ realizations (typically, $n=500$)
 by slightly modifying the procedure of \cite{hybridgeneration}:
 For each realization, we generate $\Nch'>\Nch$ (infinitely thin) independent single FENE polymer chains, each
 distributed according to $\exp(-\bLambda\colon\!\bPi^*)$, where $\bPi^*=\bPi/\Nch$ is the
 tensor of gyration of the single chain.
 Next, the diameter of chains is successively increased, and overlapping chains selectively removed.
 With this method, we generate a polymer melt at the desired density,
 where the anisotropy generated by $\bLambda$ remains preserved.
 Subsequently, Maxwellian distributed
 velocities are assigned, in agreement with (\ref{newrhoz}).
 For \QQQs one chooses a symplectic integrator (we have used a
 velocity-Verlet algorithm) to perform microcanonical equilibrium MD
 based on the microscopic Hamiltonian $H(\z)$. We calculate and store trajectories $\z(t)$
 during a short time interval, $t\in[0,\taus]$, which is small enough to not significantly alter $\x$
 during the course of the MD. For polymeric systems, the gyration tensor will relax towards equilibrium
 on a time scale $\tau$ which is known to be huge compared with the Lennard-Jones time unit,
 $\tau = 0.39\,(1+N/78)N^2$ from \cite{mkbook}
 for melts under study, i.e., $\tau\approx 200$ for $N=20$.
 As we carefully investigated, results are (as they should for proper choice of $\taus$)
 insensitive on $\taus$ in the regime $\tau/\taus\in[5,50]$. See also Fig.~\ref{coarse_grain_2008_FigM}.
 We use $\taus=\tau/30\ll \tau$, and $N\in\{10,20,30\}$ for results to be presented.
Notice, that the MD simulation time is thus very short compared to conventional
nonequilibrium MD (NEMD) at (the problematic) low field strengths (flow rates), where
simulation times large compared with the inverse rate ($\dot{\gamma}^{-1}$) are required.
 \QQQQs With the $n$ sets of phase space trajectories $\z(t)$ at hand, one inserts them into the definition
 of the slow variable $\bPi(\z(t))$, and then
 evaluates the friction term ${\bf M}$ (in our case a $4\times4$ matrix) from (\ref{GreenKubo}),
 with $\triangle_{\taus}\bbf{\Pi}(\z) \equiv \bbf{\Pi}^*(\z(\taus)) - \bbf{\Pi}^*(\z(0))$.
 The average in (\ref{GreenKubo})
 is evaluated as an arithmetic mean over the $n$ independent trajectories,
 e.g., ${\bf M}=(1/n)\sum_i \bbf{\cal M}_{(i)}$, where we denote
 the partial contribution from trajectory $i\in\{1,..,n\}$ by a bracketed subscript.
 The number of samples $n$ has to be chosen large enough to calculate ${\bf M}$
 sufficiently accurate.
 In our case, several components of ${\bf M}$ should vanish by symmetry consideration,
 and one can choose $n$ as large as to ensure these components vanish within
 statistical uncertainty. Notice further, that ${\bf M}$ possesses basic symmetries
 such as $M_{\alpha\beta\mu\nu}=M_{\mu\nu\alpha\beta}=M_{\beta\alpha\mu\nu}$ for
 arbitrary choices of indices
 because $\bPi$ is symmetric.
 \QQQQQs Repeating the
 procedure \Qn--\QQQQQn--\Qn--.. for each $\dot{\gamma}$ until convergence
 can be replaced by an efficient reweighting scheme.
 This scheme relies on the smallness of the change of increment $\Delta a$, which comes together with
 moderate changes of the distribution function $\rho$.
 To this end we use Broyden's method with standard settings
 \cite{recipes2006}
 which does not require the Jacobian matrix,
 to solve the nonlinear system
 \be
 {\bf 0} =
   \sum_{i=1}^n \left[ {\bf C}_{i} +  \kB \bbf{\cal M}_{(i)}\!\colon\!\DeltaLambda \right]
    w_i, \; w_i \equiv \frac{e^{-\DeltaLambda:\bPi_{(i)}} }{\sum_j e^{-\DeltaLambda:\bPi_{(j)}}}
 \label{newMeqabb}
 \ee
 for (matrix) $\DeltaLambda$, with mismatch
 ${\bf C}_{i} \equiv \dot{\x}_{\rm rev}(\bPi_{(i)},\bkappa) +
 \kB \bbf{\cal M}_{(i)}\!\colon\!\bLambda$,
 cf.\ Eqs.~(\ref{newrhoz}), (\ref{xdot}).
 For example, in a shear flow,
 (\ref{newMeqabb}) stands for six equations and six unknowns.
 With the solution $\DeltaLambda$ of (\ref{newMeqabb}) at hand, we directly
 calculate the reweighted slow variables and friction matrix,
 $\x=\sum_i w_i \bPi_{(i)}$,
 ${\bf M}=\sum_i w_i \bbf{\cal M}_{(i)}$,
 as well as updated Lagrange multipliers,
 $\bLambda\rightarrow \bLambda+\DeltaLambda$.
A justification of the reweighting scheme is given in Appendix \ref{reweight.app}.
 Finally, we increase the control parameter $\log \dot{\gamma}\rightarrow \log \dot{\gamma} + \Delta a$,
 and start over with step \Qns of the procedure, until we have
 swept through the control parameter space.

By then, we have recorded
 consistent sets $\x$, ${\bf M}$, as well as $\bLambda$ for the whole range of parameters $\dot{\gamma}$.
 That is, we have obtained $\bLambda(\x)$ and ${\bf M}(\x)$ and therefore established the coarse-grained model
 (\ref{xdot}) for particular parameterized path $\x(\dot{\gamma})$.
 By choosing the control parameters appropriately, our approach uses paths to explore those regions in state and
 parameter space that correspond to driven nonequilibrium situations of interest.
 For the system under study, the quantity $\x(\dot{\gamma})$ itself is experimentally accessible by means
 of small angle neutron scattering \cite{mkbook}.
 Other particularly interesting material functions are flow curves, i.e.,
 stress tensor $\stress$ as function of the control parameter $\dot{\gamma}$.
 The macroscopic expression for the polymer contribution to the stress tensor
\begin{equation} \label{stressxL}
 \stress = -2 n_p \kB T\, \x \cdot\!\bLambda,
\end{equation}
 where $n_p$ is the polymer concentration,
 follows from both, nonequilibrium thermodynamics \cite{hcobook2},
 and by evaluating the microscopic expression for the stress tensor
 in the ensemble (\ref{newrhoz}), see Appendix \ref{stresstensor.app}.
A more detailed discussion of the stress tensor within this context
is given in \cite{pi_unentangled}.

Before presenting results obtained with the proposed method, we
briefly comment on the time- and length-scales involved, already
alluded to in the introduction. The original, microscopic model has
as characteristic length scale the bead diameter $\sigma$ and
reference time $\tau_{\rm LJ}=[m\sigma^2/\epsilon]^{1/2}$, where $m$
is the mass and $\epsilon$ the characteristic Lennard-Jones
interaction energy. On the coarse-grained level, the characteristic
length scale is the radius of gyration, $R_g\approx \sigma N^{1/2}$.
The corresponding time scale estimated from the Rouse model
\cite{Rouse} is $\tau_R=\zeta (N\sigma)^2/[3\pi^2\kB T]$, where
$\zeta$ is the bead friction coefficient. Therefore, the bridging of
length scale $R_g/\sigma=N^{1/2}$ is associated with a bridging of
time scales $\tau_R/\tau_{\rm LJ}=cN^2$, where
$c=5/(16\pi^{3/2})[\zeta/\zeta_0][\epsilon/\kB T]^{1/2}$ with 
$\zeta_0=3\pi/(16\sigma)[\pi m\kB T]^{1/2}$ the friction coefficient 
of a hard sphere gas.

\section{Results}
\begin{figure}[tb]
\begin{center}
\includegraphics[width=9cm, angle=0]{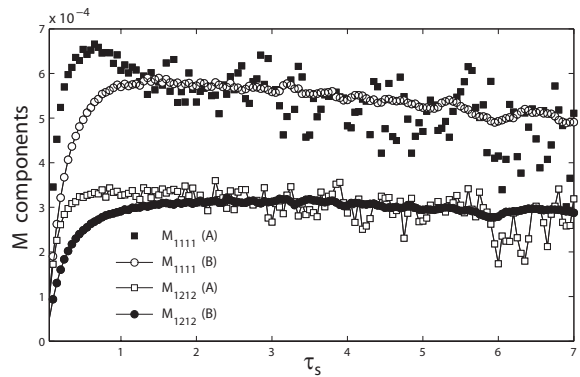}
\end{center}
\caption{Different components of the friction matrix $\M$ as a function of the
separating time $\taus$ obtained in step \QQQs of our procedure. Solid and open symbols correspond
to the integral formula mentioned in the text and Eq.~(\ref{GreenKubo}), respectively.
Results correspond to a chain length of $N=30$ and a planar shear flow with dimensionless shear rate
$\dot{\gamma}=0.00036$.}
\label{coarse_grain_2008_FigM}
\end{figure}

Figure \ref{coarse_grain_2008_FigM} shows different components of
the $\M$ matrix (\ref{GreenKubo}) as a function of the separating time $\taus$. As
mentioned before, the results for $\M$ are to a good approximation
independent of the precise value of $\taus$ in a broad range
$\taus\in [5,50]$ which is significant smaller than the polymer relaxation time 
$\tau$ ($\tau\approx 200$ for $N=20$). Furthermore, the comparison in
Fig.~\ref{coarse_grain_2008_FigM} shows that the simplified formula
(\ref{GreenKubo}) approximates the more accurate integral formula
\cite{hcobook2} $\M = \frac{1}{\kB}\int_0^{\taus}\!{\rm
d}t\avee{\dot{\bPi}(t)\dot{\bPi}(0)}$ quite well.

Having established the thermodynamic building blocks $\bLambda(\x)$ and ${\bf M}(\x)$, we can use the
evolution equations (\ref{xdot}) to study time-dependent flows.
 We have calculated
 transient dynamics in startup of steady shear flow,
 or storage and loss moduli $G'$ and $G''$ as function of frequency $\omega$
 upon using an oscillating control parameter $\dot{\gamma}\propto \sin\omega t$ (graphs not shown).
 We note that, due to our choice of the parameterization $\x(\dot{\gamma})$, the transient dynamics
 $\x(t)$ is readily calculated as long as we do not leave the known subspace $\{\x(\dot{\gamma})\}$.
 Otherwise, interpolation and extrapolation methods are needed for parameterizing the missing regions
 in $\x$-space.

There are several options to test the range of validity
 of the coarse-grained model.
 As an internal consistency check, we recommend comparing the macroscopic expression for the stress tensor
 Eq.~(\ref{stressxL}) with the standard microscopic (virial) expression, Eq.~(\ref{stress_def}).
Both are available during the course
 of the simulation. We have verified that the two expressions for $\stress$ agree with each other for the
 range of flow rates considered.
Under strong flow conditions and beyond the scope of the present
study, higher order modes and kinetic contributions to the stress
tensor tend to become increasingly important and need to be included
suitably in $\x$, cf.~\cite{ilgcanonical1}.

\begin{figure}[tb]
\begin{center}
\includegraphics[width=9cm, angle=0]{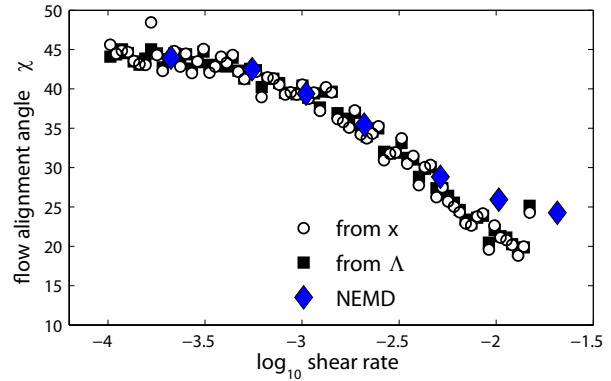}
\end{center}
\caption{(color online). Flow alignment angle $\chi$ calculated from gyration tensor $\x$ 
(open circles) and $\bLambda$ (filled squares) as a
function of the logarithm of the shear rate $\dot{\gamma}$ for planar shear
flow (chain length $N=30$). Diamonds correspond to NEMD reference results 
taken from \cite{mkbook} obtained under the same conditions.} \label{coarse_grain_2008_Figalign}
\end{figure}

\begin{figure}[tb]
\begin{center}
\includegraphics[width=9cm, angle=0]{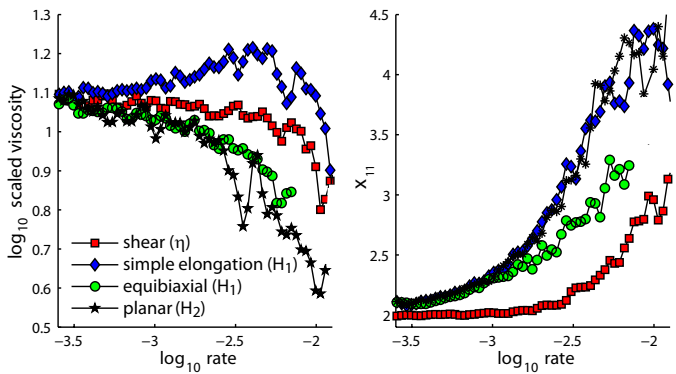}
\end{center}
\caption{(color online). {\bf (a)} We follow the notation employed
in \cite{equibiaxial} where the transposed flow gradient is written
as $\nabla{\bf v} = \dot{\epsilon}\left[\e_1\e_1 + m \e_2\e_2  -
(1+m)\e_3\e_3 \right] + \dot{\gamma}\e_1\e_2$, with shear rate
$\dot{\gamma}$, elongation rate $\dot{\epsilon}$, special cases
$m=-0.5$ (simple), $0$ (planar), $+0.5$ (elliptical), and $1$
(equibiaxial elongation) when $\dot{\gamma}=0$, and simple shear,
when $\dot{\epsilon}=0$. Besides shear viscosity $\eta$, the graph
shows the properly (cf.~text part and \cite{equibiaxial,daivis})
scaled viscosities $H_1\equiv\eta_1/[2(2+m)]$ and
$H_2\equiv\eta_2/[2(1+2m)]$ vs.~flow rate for $N=20$, where $\eta_1
\equiv (\sigma_{11}-\sigma_{33})/\dot{\epsilon}$ and $\eta_2 \equiv
(\sigma_{22}-\sigma_{33})/\dot{\epsilon}$. {\bf (b)} Maximum
component of the gyration tensor $\x_{11}$ for the same types of
flow, vs.~flow rate ($N=20$).} \label{coarse_grain_2008_Fig2}
\end{figure}

\begin{figure}[tb]
\begin{center}
\includegraphics[width=9cm, angle=0]{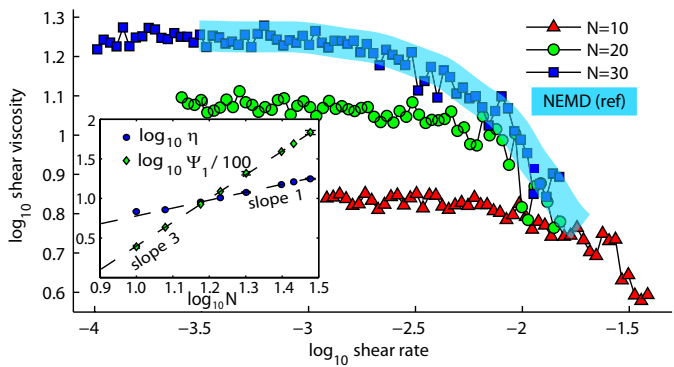}
\end{center}
\caption{(color online). Polymer contribution to non-newtonian shear
viscosity vs.~shear rate for various molecular weights. Exemplarily,
reference results obtained via direct NEMD simulation \cite{mkbook}
are shown for $N=30$. The inset shows zero-rate shear viscosity
$\eta_0$ and first viscometric function $\Psi_{1,0}$ vs.~chain
length $N$, both coinciding with data from extensive NEMD
\cite{mkbook}.} \label{coarse_grain_2008_Fig3}
\end{figure}

We apply the proposed method to
 the FENE polymer melt described above, subjected to various
 flows (results for mixed and elliptical elongational flow not shown).
 For the case of simple shear, Fig.~\ref{coarse_grain_2008_Fig1} shows the shear rate dependence of the chosen slow
 variable $\x$ (tensor of gyration in our case) and the corresponding Lagrange multiplier $\bLambda$.
Very good agreement of $\x$ with NEMD reference results is obtained.
 As a further consistency check, we have verified, that the basic identity
 $(x_{11}-x_{22})x_{12}^{-1}=(\Lambda_{11}-\Lambda_{22})\Lambda_{12}^{-1}$, derived from
 Eq.~(\ref{xdot}) \cite{vlasisPRL2007}
 using our choice for $\x$ and $\bkappa$, holds within error margins. 
This quantity is related to the flow alignment angle $\chi$ by 
$(x_{11}-x_{22})x_{12}^{-1}=2\cot(2\chi)$. Therefore, we show in 
Fig.~\ref{coarse_grain_2008_Figalign} the alignment angle $\chi$ calculated from $\x$ as well 
as from $\bLambda$. 
The very good agreement between those values shows the intrinsic consistency of our results. 
Furthermore, our results are also in good agreement with standard NEMD simulation also displayed in 
Fig.~\ref{coarse_grain_2008_Figalign} for planar shear flow with $N=30$ \cite{mkbook}. 
 Figure \ref{coarse_grain_2008_Fig2}a shows shear and extensional viscosities for different
 flow conditions. Our results confirm expectations from
 a retarded motion expansion analysis for a comparable system, studied via extensive NEMD in \cite{daivis}.
In particular, Fig.~\ref{coarse_grain_2008_Fig2}a shows that the scaled viscosities all
superimpose for vanishing flow rates, in agreement with predictions from linear viscoelasticity theory.
Also in agreement with previous results, the viscosity in simple elongation exhibits a maximum around 
a dimensionless rate of order unity, while in planar and equibiaxial elongation as well as in planar shear flow 
the viscosity decreases monotonically with flow rate \cite{equibiaxial,daivis}. 
 The corresponding $x_{11}$-components of the gyration tensor,
 which characterize the polymer stretch,
 are plotted in Fig.~\ref{coarse_grain_2008_Fig2}b.
We observe that polymer stretching is much more pronounced for planar and equibiaxial elongation compared to  
in planar shear flow. 
 We have further validated the proposed algorithm for the rate ($\dot{\gamma}$) and chain length ($N$) dependence
 of the shear viscosity (see Fig.~\ref{coarse_grain_2008_Fig3}, which offers
 a quantitative comparison with available NEMD data from \cite{mkbook} for an identical system).
 Since our method does not require flow-adapted boundary conditions, we are able to include here the first
 simulation results on steady state equibiaxial elongation.
All results for the sample application, including the many beyond
the scope of this article and therefore not reported here, reproduce
available experimental findings for gyration tensor and viscosities
(shear thinning, strain hardening only in simple elongation,
alignment in shear weaker than in elongation at same flow
invariants, scaling behavior, overshoots,
cf.~\cite{larsonbook,equibiaxial,dealybook}). The results provided
in this section clearly demonstrate that the proposed simple
procedure outlined in Tab.~\ref{method.table} allows to (i) recover
known results obtained via classical approaches, (ii) study flow
geometries not accessible using alternate approaches, (iii)
calculate the friction matrix and Lagrange multiplier, i.e., the
irreversible part of the closed and low-dimensional time evolution equation 
(\ref{xdot}) for the coarse-grained variable in a
straightforward manner.

\section{Conclusions}
Using an alternating MC--MD iteration scheme,
 our approach successfully bridges the time-scale gap between microscopic and macroscopic scales
 by establishing the coarse-grained model within a nonequilibrium thermodynamics framework.
 Since only short MD simulations are needed, our method is very efficient
 (moreover, it is ideally suited for parallelization) and
 particularly allows to deal with arbitrary flow gradients, since neither
 special boundary conditions nor other constraints are needed.
To be specific, even from the viewpoint of material property
determination, our method is more efficient than standard NEMD when
$\dot{\gamma}\tau <(n_0/n)(\tau/\taus)$, where $n_0$ denotes the
number of strain units needed for the NEMD. With $\tau/\taus=30$,
$n=500$ used here, taking $n_0=10$ from \cite{daivis} and also the
time for the MC step into account (see \cite{hybridgeneration}), our
method is superior to NEMD for $\dot{\gamma}\tau\le 0.5$, and $m$
orders of magnitude faster at a value $10^m$ times smaller than that
dimensionless rate.

The presented approach is very general, but the (a-posteriori
validated) success of the coarse-graining procedure depends
crucially on the proper choice of the slow variable. 
As mentioned above in our illustrating
example, conformation tensors as slow variables for polymer melts
are clearly restricted to the unentangled regime because interchain
effects, entanglements or knots, hinder the relaxation of the
conformation tensor for high molecular weight polymers
\cite{hoy,mkbook}.
Some promising candidates for other soft matter systems are the
tensorial order parameter for liquid crystals,
 the magnetization for magnetic liquids, and the path length of the entanglement network
 for entangled polymer melts \cite{larsonbook,mkbook,christoscurropin}.
 The MC step is particularly challenging for dense polymeric systems,
 but efficient schemes exist for FENE as well as for atomistic models
 \cite{vlasisPRL2007,nikosPRL2002}.

For many complex fluids, Eq.~(\ref{newrhoz})
is known to serve as a successful starting point to derive closure relationships \cite{mkbook,ilgcanonical1}.
Therefore, our method establishes the coarse-grained model all the way from equilibrium up to the
validity of (\ref{newrhoz}) and complements standard NEMD methods, which remain often well-suited for
the less challenging regime of strong external forcing.

\section*{Acknowledgments}
 We acknowledge support through grants NMP3-CT-2005-016375
 and FP6-2004-NMP-TI-4 STRP 033339 of the European Community.

\begin{appendix}

\section{Choice of slow variables} \label{slow.app}
The proper choice of appropriate slow (collective) variables is crucial not only for the method 
proposed here but for a broad class of nonequilibrium statistical mechanics approaches based on 
projection operator techniques \cite{hcobook2}. 

In the present case of unentangled polymer melts, there is ample evidence that 
single chain conformation tensors are promising candidates for slow variables 
\cite{larsonbook,hcobook2,vlasisPRL2007}. 
Therefore, we assume the slow variable $\x$ can be decomposed into an average over 
single chain (symmetric second rank) conformation tensors, 
\begin{equation}
\x = \frac{1}{\Nch}\sum_{a=1}^{\Nch}\x^{(a)}.
\end{equation}
The latter can always be expanded in a series of Rouse modes, 
\begin{equation} \label{expandmodes}
\x^{(a)} = \sum_{p=1}^{N-1}c_p {\bf X}_p^{(a)}{\bf X}_p^{(a)}; \quad 
{\bf X}_p^{(a)} = \sum_{j=1}^{N-1}\Omega_{pj}{\bf Q}_j^{(a)},
\end{equation}
where ${\bf X}_p^{(a)}$ is the $p$-th Rouse mode of chain $a$, 
$\Omega_{pj}=\sqrt{2/N}\sin(p\pi j/N)$ an element of the Rouse matrix and 
${\bf Q}_j^{(a)}$ the connector vector of particles $j+1$ and $j$ of chain $a$ 
\cite{larsonbook}. 

As possible choice of the weights $c_p$ in Eq.~(\ref{expandmodes}), we initially implemented 
$c_p=\delta_{p1}$, i.e.~only the first Rouse mode is included. 
This choice is reasonable since the first Rouse mode is the slowest and therefore a 
natural candidate for the slow variable $\x$. 
However, the resulting model is restricted to very small deviations from equilibrium 
because there is no clear time scale separation to the higher modes which are neglected. 
In fact, the relaxation time of mode $p$ in the Rouse model is 
$\tau_p=\zeta/[8k\sin^2(p\pi/2N)]\propto \tau_1/p^2$ 
($\zeta$ and $k$ are the bead friction coefficient and the spring constant in the 
Rouse model, respectively)
and therefore the second mode relaxes only a factor four faster than the first one. 
Thus, when driven out of equilibrium like in a flow situation, several of the lowest 
Rouse modes are typically excited. 
In order to address this issue, we propose to include all Rouse modes in a single 
quantity such that the increasing relaxation times of the higher modes are 
reflected in a decreasing weight $c_p$. 
Such a choice can be motivated by the fact that,  
in a stationary flow situation, the Lagrange multiplier is proportional to the 
product of relaxation time and velocity gradient 
(see e.g.~Eq.~(8.52) in \cite{hcobook2}).  
Thus, we use a single Lagrange multiplier $\bLambda$ in the nonequilibrium ensemble 
Eq.~(\ref{newrhoz}) in order to excite all Rouse modes at the same time in a way 
that is consistent with Rouse theory. 
With $c_p=N/(\pi p)^2$ decreasing for increasing mode number $p$ as the corresponding 
relaxation times, $\x$ becomes the gyration tensor, at least to a very good approximation. 
Comparing different choices for $\x$ (gyration tensor and the second rank tensor 
formed by either the first Rouse mode or the end-to-end vector), we found the 
gyration tensor to give the most accurate results.

\section{Reweighting scheme} \label{reweight.app}
We here describe the reweighting scheme employed in the above
described algorithm. The generalized canonical distribution
(\ref{newrhoz}) is denoted by $\rho_{\Lambda}(\z)$, in order to make
explicit its dependence on the values of the Lagrange multipliers
$\bLambda$. Corresponding averages of phase space functions are
$\avee{A}_\Lambda\equiv\int\!{\rm d}\z\, A(\z)\rho_{\Lambda}(\z)$.
The analytical form of (\ref{newrhoz}) allows to relate the
distribution $\rho_{\Lambda+\delta\Lambda}(\z)$ corresponding to
different Lagrange multipliers $\bLambda+\delta\bLambda$ to
$\rho_{\Lambda}(\z)$ by
\begin{equation} \label{rho_delambda}
 \rho_{\Lambda+\delta\Lambda}(\z) = \rho_{\Lambda}(\z)\,
 \frac{e^{-\delta\bLambda:\bPi(\z)}}{\ave{e^{-\delta\bLambda:\bPi(\z)}}_\Lambda}.
\end{equation}
Therefore, also the averages of phase space functions corresponding to different
values of Lagrange multipliers are related by
\begin{equation} \label{A_delambda}
 \ave{A}_{\Lambda+\delta\Lambda} =
 \frac{\ave{Ae^{-\delta\bLambda:\bPi(\z)}}_{\Lambda}}{\ave{e^{-\delta\bLambda:\bPi(\z)}}_\Lambda}.
\end{equation}
For small deviations $\delta\Lambda$, the latter expression
simplifies to
\begin{equation} \label{A_delambdaapprox}
 \ave{A}_{\Lambda+\delta\Lambda} \approx 
\frac{\ave{A}_\Lambda - \delta\bLambda:\ave{\bPi A}_\Lambda}{1-\delta\bLambda:\ave{\bPi}_\Lambda}.
\end{equation}
Equation (\ref{A_delambda}) or (\ref{A_delambdaapprox}) for $A=\bPi$
and $A=\bbf{\cal M}$ are used in the reweighting scheme in order to
calculate the corrected values for $\x$ and $\M$, respectively, from
recorded averages. In principle, Eq.~(\ref{A_delambda}) allows to
recalculate averages of $A$ for arbitrary $\delta\bLambda$. In
practice, however, due to the finite ensemble size, such estimates
are accurate only if $\rho_\Lambda$ and
$\rho_{\Lambda+\delta\Lambda}$ have considerable overlap. This is
the case for $\delta\bLambda$ small enough such that relevant states
for averages at $\bLambda+\delta\bLambda$ are sufficiently well
sampled with $\rho_\Lambda$.

\section{Stress tensor in generalized canonical ensemble} \label{stresstensor.app}

Point of departure is the microscopic expression for the total
stress tensor \cite{larsonbook} which can be inferred from the term
of second order in the expansion of the configurational Helmholtz
free energy with respect to the strain tensor \cite{Hess97}
\begin{equation} \label{stress_def}
 \sigma_{\alpha\beta}^{\rm tot} = - \frac{1}{V}\ave{\sum_j m_j c_{j,\alpha}c_{j,\beta}}
 - \frac{1}{V}\ave{\sum_{j}r_{j,\alpha}F_{j,\beta}}.
\end{equation}
The first, kinetic contribution can be well approximated by the
ideal gas expression with $p\equiv n\kB T$. Deviations from this
expression are minor in polymer melts and show up only at extremely
high flow rates \cite{mkbook,mk_chainend}.

We further assume (i) potential forces
$F_{j,\alpha}=-\partial H/\partial r_{j,\alpha}$,
and (ii) a generalized canonical ensemble
$\rho(z)=(1/Z^\ast)\exp[-\beta H - \Nch\bLambda\colon\bPi]$,
cf.~Eq.~(\ref{newrhoz}),
where $\Nch$ denotes the number of polymer chains.

Assumptions (i) and (ii) allow us to write
\begin{eqnarray}
  F_{j,\beta}\,\rho & = &
  -\frac{\partial H}{\partial r_{j,\beta}} \rho\nonumber\\
  & = & \frac{1}{\beta}\frac{\partial}{\partial r_{j,\beta}} \rho
  + \frac{\Nch}{\beta} \Lambda_{\mu\nu}\frac{\partial \Pi_{\mu\nu}}{\partial r_{j,\beta}}\rho.
\end{eqnarray}
Inserting this into (\ref{stress_def}) gives
\begin{eqnarray} \label{stress_gen}
\sigma_{\alpha\beta}^{\rm tot} & = &
      -p \delta_{\alpha\beta}
      -\frac{1}{\beta V}\sum_j
       \underbrace{\int\!{\rm d}\z\, r_{j,\alpha} \frac{\partial}{\partial r_{j,\beta}}
       \rho(\z)}_{=0}
      \nonumber \\
&& - \frac{\Nch\Lambda_{\mu\nu}}{V\beta}
       \sum_j\int\!{\rm d}\z\, r_{j,\alpha}\frac{\partial \Pi_{\mu\nu}}{\partial r_{j,\beta}}
       \rho(\z)
\nonumber\\
& = &
      - p \delta_{\alpha\beta}
      - n_p\kB T\Lambda_{\mu\nu}
        \ave{\sum_j r_{j,\alpha}\frac{\partial \Pi_{\mu\nu}}{\partial r_{j,\beta}}}.
\end{eqnarray}

For the special case of conformation tensor models, $\bPi$ can be expressed as a
bilinear form of the particle positions. Then, we obtain from
Eq.~(\ref{stress_gen}) the final expression
\begin{equation} \label{stress_conftensor}
 \bsigma^{\rm tot} = -p {\bf 1} -2n_p\kB T\x\cdot\bLambda.
\end{equation}

Equation (\ref{stress_conftensor}) can independently be derived from
nonequilibrium thermodynamics \cite{hcobook2}. It should be noted
that Eq.~(\ref{stress_conftensor}) captures the polymer contribution
to the stress tensor as the Lagrange multipliers $\bLambda$
describe nonequilibrium polymer configurations. For short-chain
polymer melts, a ``simple fluid contribution'' has to be added in
order to account for the stress contribution to the total stress
tensor that would be present in the absence of chain connectivity
\cite{Barrat_Rouse,pi_unentangled}.

\end{appendix}


\begin{thebibliography}{99}
\bibitem{gellmann2007}
M.~Gell-Mann, J.~B.~Hartle, Phys. Rev. A {\bf 76}, 022104 (2007).

\bibitem{larsonbook}
R.~G.~Larson, {\em The structure and rheology of complex fluids}
(Oxford University Press, New York, 1999).

 \bibitem{krakoviackCGcopolym}
C.~Pierleoni, C.~Addison, J.-P.~Hansen, V.~Krakoviack,
Phys. Rev. Lett. {\bf 96}, 128302 (2006).

 \bibitem{kleinCGreview}
S.~O.~Nielsen, C.~F.~Lopez, G.~Srinivas, M.~L.~Klein, J. Phys.:
Condens. Mat. {\bf 16}, R481 (2004).


 \bibitem{coarsegrained2000}
J.~Baschnagel {\em et al.},
Adv. Polym. Sci. {\bf 152}, 41 (2000).

\bibitem{Kremer_reviewMultiscale}
M.~Praprotnik, L.~D.~Site, K.~Kremer, Ann.~Rev. Phys. Chem. {\bf
59}, 545 (2008).

\bibitem{guenza_reviewCG}
M.~G.~Guenza, J.~Phys.: Condens.~Mat.~{\bf 20}, 033101 (2008).

\bibitem{curtarolo}
S.~Curtarolo, G.~Ceder,
Phys. Rev. Lett. {\bf 88}, 255504 (2002).

\bibitem{Larson_CGchain}
R.~G.~Larson,
Mol.~Phys. {\bf 102}, 341 (2004).

\bibitem{Briels_CGchain}
R.~L.~C.~Akkermans, W.~J.~Briels,
J.~Chem.~Phys. {\bf 114}, 1020 (2001).

\bibitem{BrielsPadding}
J.~T. Padding, W.~J.~Briels,
J.~Chem.~Phys. {\bf 117}, 925 (2002).

\bibitem{hco_lessons}
H.~C.~\"Ottinger, MRS Bull. {\bf 32}, 936 (2007).

 \bibitem{hcobook2}
H.~C.~\"Ottinger, {\em Beyond Equilibrium Thermodynamics } (Wiley,
Hoboken NJ, 2005).

 \bibitem{mkbook}
M.~Kr\"oger, W.~Loose, S.~Hess, J. Rheol. {\bf 37}, 1057 (1993);
Phys. Rev. Lett. {\bf 85}, 1128 (2000); M. Kr\"oger, {\em Models for
Polymeric and Anisotropic Liquids } (Springer, Berlin, 2005).

\bibitem{daivis}
P.~J.~Daivis, M.~L.~Matin, B.~D.~Todd, J. Non-Newton. Fluid Mech.
{\bf 147}, 35 (2007); ibid.~{\bf 111}, 1 (2003).

\bibitem{hoy} K.~Kremer, G.~S. Grest, J. Chem. Phys. {\bf 92}, 5057 (1990);
R.~S. Hoy, G.~S.~Grest, Macromolecules {\bf 40}, 8389 (2007).

\bibitem{units} Reduced units - online interactive tool permanently available at
http://www.complexfluids.ethz.ch/units

\bibitem{toddreview}
B.~D.~Todd, P.~J.~Daivis, Mol. Simul. {\bf 33}, 189 (2007); Phys.
Rev. Lett. {\bf 81}, 1118 (1998).

\bibitem{ilgcanonical1}
P.~Ilg, I.~V.~Karlin, H.~C.~\"Ottinger, Physica A {\bf 315}, 367
(2002).

 \bibitem{hco1989}
H.~C.~\"Ottinger, Y.~Rabin, J.~Rheol.~{\bf 33}, 725 (1989).

 \bibitem{pi_unentangled}
P.~Ilg,
Physica A {\bf 387}, 6484 (2008).

 \bibitem{hcoprojectors}
H.~C.~\"Ottinger, Phys. Rev. E {\bf 57}, 1416 (1998).

 \bibitem{hybridgeneration}
M.~Kr\"oger,
Comput. Phys. Commun. {\bf 118}, 278 (1999).

 \bibitem{recipes2006}
W.~H.~Press, S.~A.~Teukolsky, W.~T.~Vetterling, B.~P.~Flannery, {\em
Numerical Recipes. The art of scientific computing, 3rd.~Ed. }
(Cambridge University Press, NY, 2006).

 \bibitem{Rouse} P.~E. Rouse, J. Chem. Phys. {\bf 21}, 1272 (1953).

 \bibitem{vlasisPRL2007}
C.~Baig, V.~G.~Mavrantzas, Phys. Rev. Lett. {\bf 99}, 257801 (2007).

 \bibitem{equibiaxial}
P.~Hachmann, J.~Meissner,
J. Rheol. {\bf 47}, 989 (2003).

 \bibitem{dealybook}
J.~M.~Dealy, R.~G.~Larson, {\em Structure and Rheology of Molten
Polymers } (Hanser, Munich, 2006).

 \bibitem{christoscurropin}
C.~Tzoumanekas, D.~N.~Theodorou, Curr. Opin. Solid State Mater. Sci.
{\bf 10}, 61 (2006).

 \bibitem{nikosPRL2002}
N.~Ch.~Karayiannis, V.~G.~Mavrantzas, D.~N.~Theodorou, Phys. Rev.
Lett. {\bf 88}, 105503 (2002).

\bibitem{Hess97} S. Hess, M. Kr\"oger, W.~G. Hoover,
Physica A {\bf 239}, 449 (1997).

\bibitem{mk_chainend}
M.~Kr\"oger, S.~Hess, Physica A {\bf 195}, 336 (1993).

\bibitem{Barrat_Rouse}
 M.~Vladkov, J.-L.~Barrat,
 Macromol. Theory Simul. {\bf 15}, 252 (2006).

\end{thebibliography}
\end{document}